\documentclass[10pt,twocolumn,letterpaper]{article}

\usepackage{cvpr}
\usepackage{times}
\usepackage{epsfig} 
\usepackage{graphicx}
\usepackage{amsmath}
\usepackage{amssymb}

\usepackage{multirow}
\usepackage{caption} %
\usepackage{paralist} %
\usepackage{booktabs} %

\long\def\ignorethis#1{}

\renewcommand{\etal}{et~al\mbox{.}}

\newcommand{\heading}[1]{\vspace{1mm}\noindent\textbf{#1}}

\newcommand{\figref}[1]{Figure~\ref{fig:#1}}%
\newcommand{\tabref}[1]{Table~\ref{tab:#1}} %
\renewcommand{\paragraph}[1]{\vspace{1mm} \noindent\textbf{#1}}

\newcommand{\tabmargin}{\vspace{-1mm}}
\newcommand{\figmargin}{\vspace{-2mm}}

\newcommand{\ignore}[1]{}   %

\def\PE{\Phi}

\def\exposure{t}
\def\pipeline{\PE}
\def\invpipeline{\pipeline^{-1}}

\def\hdr{H}
\def\ldr{L}
\def\cliphdr{I_c}
\def\nonlinearhdr{I_n}
\def\outputhdr{\hat{\hdr}}

\def\irradiance{E}
\def\mask{\alpha}

\def\deqout{\hat{I}_{\text{deq}}}
\def\linout{\hat{I}_{\text{lin}}}
\def\halout{\outputhdr}
\def\reout{\hat{H}_{\text{ref}}}

\def\crf{\mathcal{F}}
\def\invcrf{\mathcal{G}}

\def\invcrfvector{\mathbf{g}}
\def\invcrfscalar{g}
\def\invcrfder{\invcrfscalar'}
\def\norminvcrfder{\tilde{\invcrfscalar}'}
\def\norminvcrfscalar{\tilde{\invcrfscalar}}
\def\norminvcrfvector{\tilde{\invcrfvector}}

\def\ltwo{\ell_2}

\def\logltwo{\log-\ltwo}

\def\deqloss{\mathcal{L}_{\text{deq}}}
\def\linloss{\mathcal{L}_{\text{lin}}}
\def\halloss{\mathcal{L}_{\text{hal}}}
\def\crfloss{\mathcal{L}_{\text{crf}}}
\def\perceptualloss{\mathcal{L}_{\text{p}}}
\def\tvloss{\mathcal{L}_{\text{tv}}}
\def\totalloss{\mathcal{L}_{\text{total}}}

\def\deqlossweight{\lambda_{\text{deq}}}
\def\linlossweight{\lambda_{\text{lin}}}
\def\crflossweight{\lambda_{\text{crf}}}
\def\hallossweight{\lambda_{\text{hal}}}
\def\perceptuallossweight{\lambda_{\text{p}}}
\def\tvlossweight{\lambda_{\text{tv}}}

\newcommand{\clipping}[1]{\mathcal{C}(#1)}
\newcommand{\crfmapping}[1]{\crf(#1)}
\newcommand{\invcrfmapping}[1]{\invcrf(#1)}
\newcommand{\quantizing}[1]{\mathcal{Q}(#1)}
\newcommand{\applypipeline}[1]{\pipeline(#1)}
\newcommand{\normtwo}[1]{\| #1 \|_2^2}

\newcommand{\halnet}[1]{\mathcal{C}^{-1}(#1)}

\usepackage[numbers, square]{natbib}
\usepackage{bm}
\usepackage{caption} 
\usepackage[pagebackref=true,breaklinks=true,letterpaper=true,colorlinks,bookmarks=false,citecolor=blue,linkcolor=blue]{hyperref}

\cvprfinalcopy %

\def\cvprPaperID{0942} %

\ifcvprfinal\fi
\begin{document}

\setlength{\abovedisplayskip}{5pt}
\setlength{\belowdisplayskip}{5pt}

\title{Single-Image HDR Reconstruction by Learning to Reverse the Camera Pipeline}

\author{
Yu-Lun Liu$^{1,2*}$\quad\quad
Wei-Sheng Lai$^{3*}$\quad\quad
Yu-Sheng Chen$^{1}$\quad\quad
Yi-Lung Kao$^{1}$\\
Ming-Hsuan Yang$^{3,4}$ \quad\quad
Yung-Yu Chuang$^{1}$\quad\quad
Jia-Bin Huang$^{5}$\\
\vspace{1mm}
$^{1}$National Taiwan University \quad
$^{2}$MediaTek Inc. \quad
$^{3}$Google \quad
$^{4}$UC Merced \quad
$^{5}$Virginia Tech
 \\
{\small\url{https://www.cmlab.csie.ntu.edu.tw/~yulunliu/SingleHDR}}
}

\twocolumn[{%
\renewcommand\twocolumn[1][]{#1}%
\vspace{-5mm}
\maketitle
\vspace{-5mm}
\begin{center}
    \captionsetup{type=figure}
	\footnotesize
    \begin{minipage}[c]{0.02\textwidth}
    \rotatebox[origin = c]{90}{Input LDR images}
    \end{minipage}
    \begin{minipage}[c]{0.975\textwidth}
    \centering
        \includegraphics[width=0.247\textwidth]{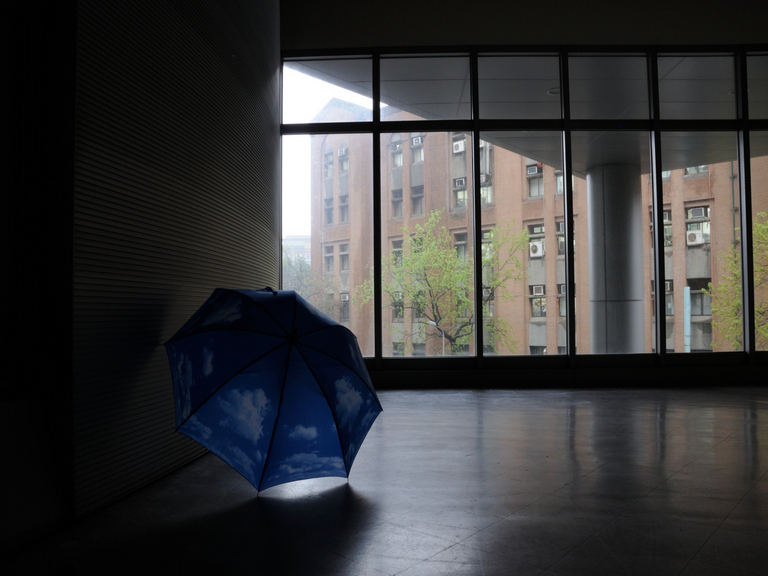} \hfill
        \includegraphics[width=0.247\textwidth]{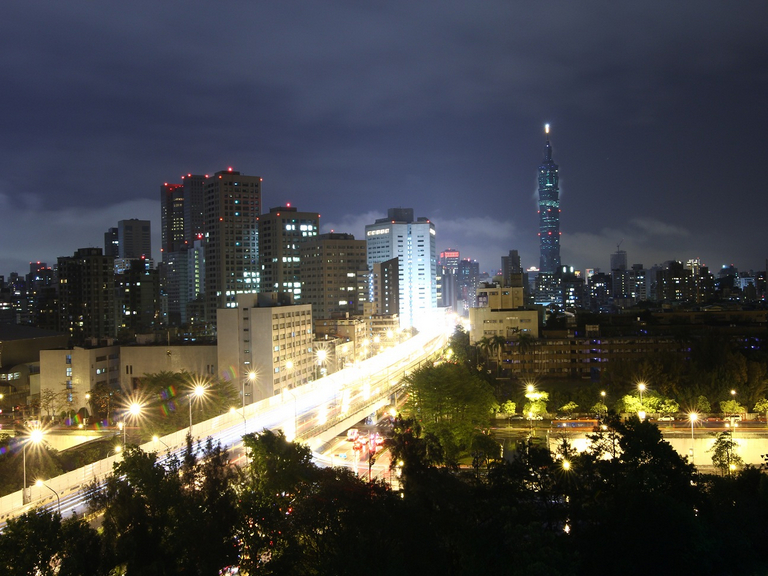} \hfill
        \includegraphics[width=0.247\textwidth]{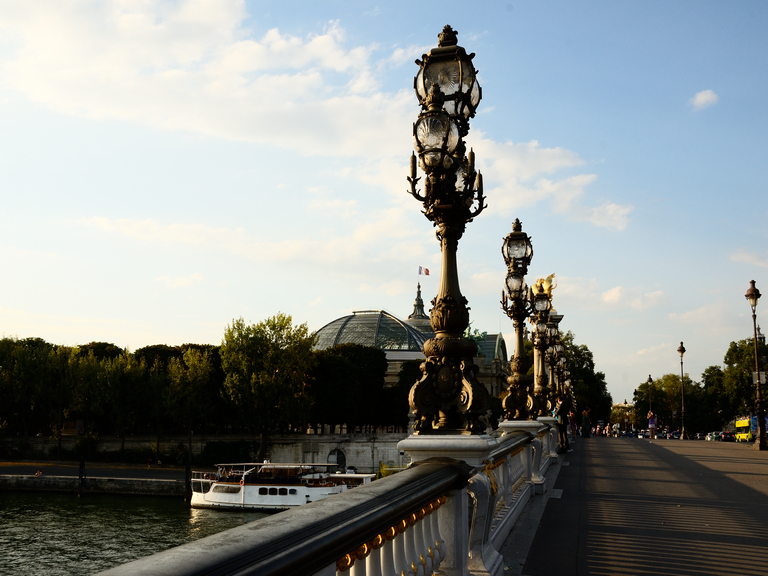} \hfill
        \includegraphics[width=0.247\textwidth]{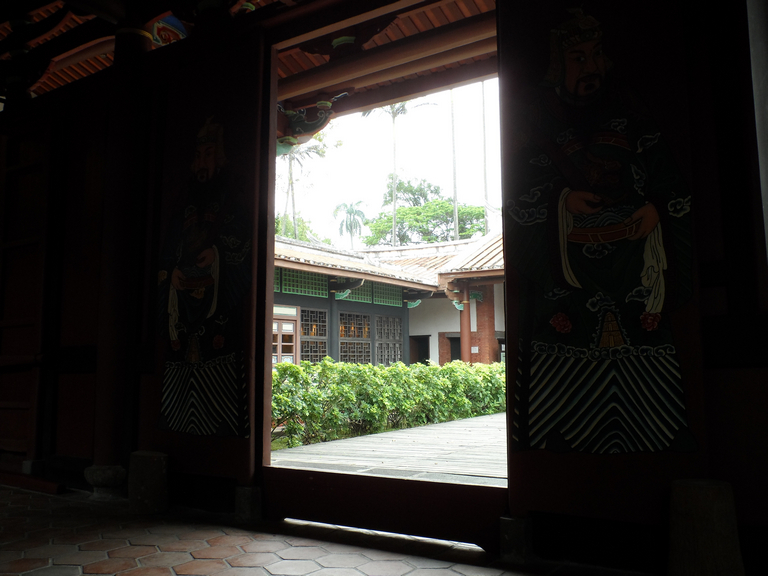}
    \end{minipage}
    \\
    \begin{minipage}[c]{0.02\textwidth}
    \rotatebox[origin = c]{90}{Our results}
    \end{minipage}
    \begin{minipage}[c]{0.975\textwidth}
    \centering
        \includegraphics[width=0.247\textwidth]{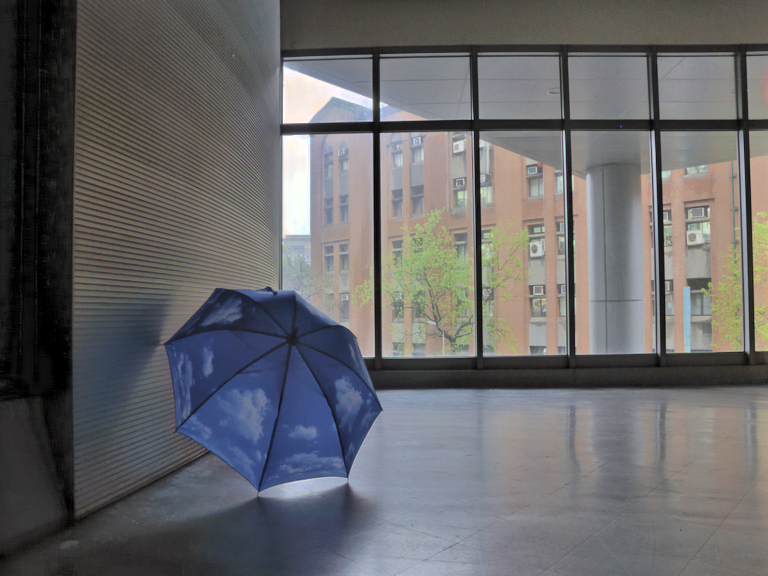} \hfill
        \includegraphics[width=0.247\textwidth]{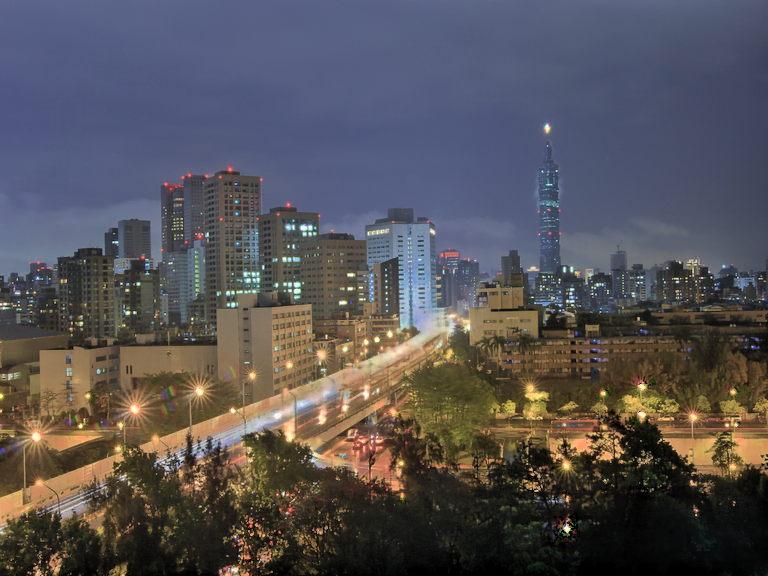} \hfill
        \includegraphics[width=0.247\textwidth]{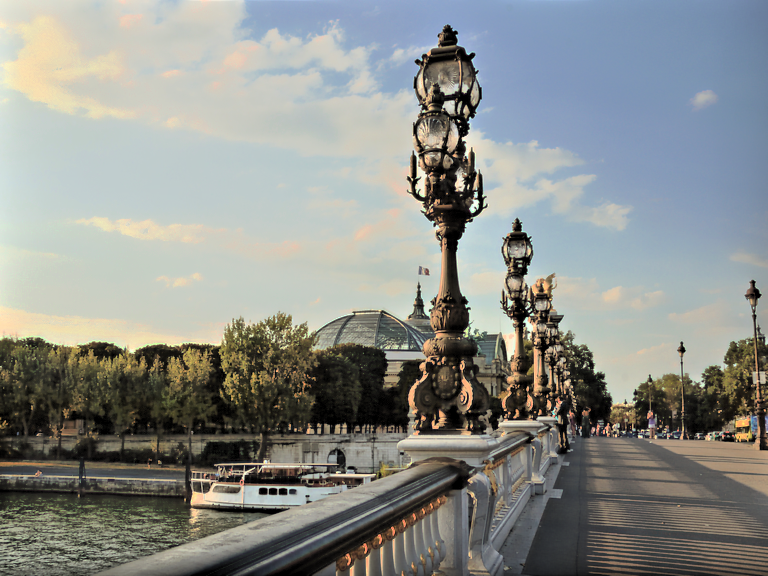} \hfill           
        \includegraphics[width=0.247\textwidth]{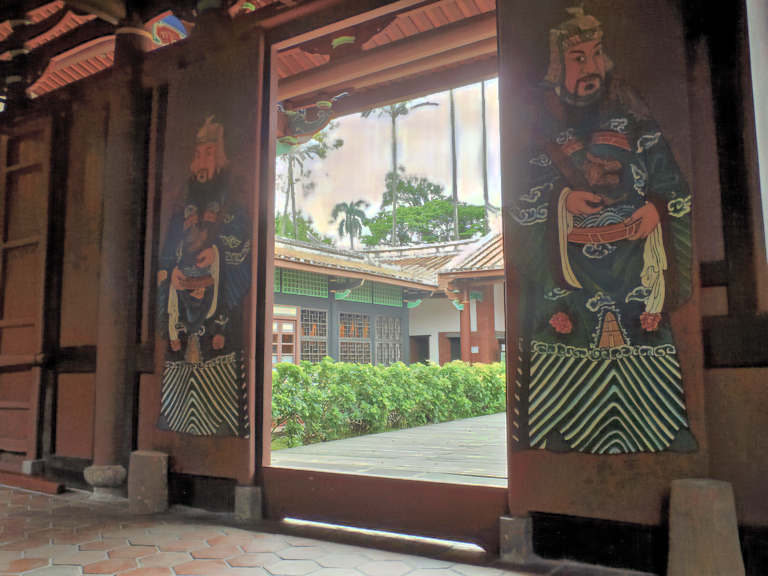}
    \end{minipage}
    \figmargin
    \caption{
        \textbf{HDR reconstruction from a single LDR image.}
        Our method recovers missing details for both backlit and over-exposed regions of real-world images by learning to reverse the camera pipeline.
        Note that the input LDR images are captured by different real cameras, and all reconstructed HDR images have been tone-mapped by~\cite{liang2018hybrid} for display.
    } 
    \label{fig:teaser}
\end{center}
}]

\newcommand\blfootnote[1]{%
  \begingroup
  \renewcommand\thefootnote{}\footnote{#1}%
  \addtocounter{footnote}{-1}%
  \endgroup
}

\blfootnote{*Indicates equal contribution.}

\vspace{-4mm}
\begin{abstract}
\vspace{-2mm}

Recovering a high dynamic range (HDR) image from a single low dynamic range (LDR) input image is challenging due to missing details in under-/over-exposed regions caused by quantization and saturation of camera sensors.
In contrast to existing learning-based methods, our core idea is to incorporate the domain knowledge of the LDR image formation pipeline into our model.
We model the HDR-to-LDR image formation pipeline as the (1) dynamic range clipping, (2) non-linear mapping from a camera response function, and (3) quantization.
We then propose to learn three specialized CNNs to reverse these steps.
By decomposing the problem into specific sub-tasks, we impose effective physical constraints to facilitate the training of individual sub-networks.
Finally, we jointly fine-tune the entire model end-to-end to reduce error accumulation.
With extensive quantitative and qualitative experiments on diverse image datasets, we demonstrate that the proposed method performs favorably against state-of-the-art single-image HDR reconstruction algorithms.
\vspace{-3mm}
\end{abstract}

\vspace{-13mm}
\section{Introduction}
\label{sec:intro}

HDR images are capable of capturing rich real-world scene appearances including lighting, contrast, and details.
Consumer-grade digital cameras, however, can only capture images within a limited dynamic range due to sensor constraints.
The most common approach to generate HDR images is to merge multiple LDR images captured with different exposures~\cite{debevec1997recovering}.
Such a technique performs well on static scenes but often suffers from ghosting artifacts on dynamic scenes or hand-held cameras.
Furthermore, capturing multiple images of the same scene may not always be feasible (e.g., existing LDR images on the Internet).

Single-image HDR reconstruction aims to recover an HDR image from a single LDR input.
The problem is challenging due to the missing information in under-/over-exposed regions.
Recently, several methods~\cite{eilertsen2017hdrcnn,endo2017drtmo,marnerides2018expandnet,yang-cvpr18-DRHT,zhang2017learning} have been developed to reconstruct an HDR image from a given LDR input using deep convolutional neural networks (CNNs). 
However, learning a direct LDR-to-HDR mapping is difficult as the variation of HDR pixels (32-bit) is significantly higher than that of LDR pixels (8-bit).
Recent methods address this challenge either by focusing on recovering the over-exposed regions~\cite{eilertsen2017hdrcnn} or synthesizing several up-/down-exposed LDR images and fusing them to produce an HDR image~\cite{endo2017drtmo}.
The artifacts induced by quantization and inaccurate camera response functions (CRFs) are, however, only \emph{implicitly} addressed through learning.

In this work, we incorporate the domain knowledge of the LDR image formation pipeline to design our model.
We model the image formation with the following steps~\cite{debevec1997recovering}: (1) dynamic range clipping, (2) non-linear mapping with a CRF, and (3) quantization.
Instead of learning a direct LDR-to-HDR mapping using a generic network, our core idea is to decompose the single-image HDR reconstruction problem into three sub-tasks: i) dequantization, ii) linearization, and iii) hallucination, and develop three deep networks to specifically tackle each of the tasks.
First, given an input LDR image, we apply a Dequantization-Net to restore the missing details caused by quantization and reduce the visual artifacts in the under-exposed regions (e.g., banding artifacts).
Second, we estimate an inverse CRF with a Linearization-Net and convert the non-linear LDR image to a \emph{linear} image (i.e., scene irradiance).
Building upon the empirical model of CRFs~\cite{grossberg2003space}, our Linearization-Net leverages the additional cues from edges, the intensity histogram and a monotonically increasing constraint to estimate more accurate CRFs.
Third, we predict the missing content in the over-exposed regions with a Hallucination-Net.
To handle other complicated operations (e.g., lens shading correction, sharpening) in modern camera pipelines that we do not model, we use a Refinement-Net and jointly fine-tune the whole model end-to-end to reduce error accumulation and improve the generalization ability to real input images.

By explicitly modeling the \emph{inverse} functions of the LDR image formation pipeline, we significantly reduce the difficulty of training one single network for reconstructing HDR images.
We evaluate the effectiveness of our method on four datasets and real-world LDR images.
Extensive quantitative and qualitative evaluations, as well as the user study, demonstrate that our model performs favorably against the state-of-the-art single-image HDR reconstruction methods.
\figref{teaser} shows our method recovers visually pleasing results with faithful details.
Our contributions are three-fold:
\begin{compactitem}
\item We tackle the single-image HDR reconstruction problem by reversing image formation pipeline, including the dequantization, linearization, and hallucination.
\item We introduce specific physical constraints, features, and loss functions for training each individual network.
\item We collect two HDR image datasets, one with synthetic LDR images and the other with real LDR images, for training and evaluation.
We show that our method performs favorably against the state-of-the-art methods in terms of the HDR-VDP-2 scores and visual quality on the collected and existing datasets. 
\end{compactitem}

\section{Related Work}
\label{sec:related}

\heading{Multi-image HDR reconstruction.}
The most common technique for creating HDR images is to fuse a stack of bracketed exposure LDR images~\cite{debevec1997recovering, Picard95onbeing}.
To handle dynamic scenes, image alignment and post-processing are required to minimize artifacts~\cite{khan2006ghost, mangiat2010high, srikantha2012ghost}.
Recent methods apply CNNs to fuse multiple flow-aligned LDR images~\cite{kalantari2017hdr} or unaligned LDR images~\cite{wu2018deep}.
In contrast, we focus on reconstructing an HDR image from a \emph{single} LDR image.

\heading{Single-image HDR reconstruction.}
Single-image HDR reconstruction does not suffer from ghosting artifacts but is significantly more challenging than the multi-exposure counterpart.
Early approaches estimate the density of light sources to expand the dynamic range~\cite{akyuz2007hdr, banterle2009itmo, banterle2006itmo, banterle2008itmo, banterle2007itmo} or apply the cross-bilateral filter to enhance the input LDR images~\cite{huo2014physiological, kovaleski2014high}.
With the advances of deep CNNs~\cite{he2016deep, VGG}, several methods have been developed to learn a direct LDR-to-HDR mapping~\cite{marnerides2018expandnet,yang-cvpr18-DRHT,zhang2017learning}.
Eilertsen~\etal~\cite{eilertsen2017hdrcnn} 
propose the HDRCNN method that focuses on recovering missing details in the over-exposed regions while ignoring the quantization artifacts in the under-exposed areas.
In addition, a fixed inverse CRF is applied, which may not be applicable to images captured from different cameras.
Instead of learning a direct LDR-to-HDR mapping, some recent methods~\cite{endo2017drtmo,lee2018deep} learn to synthesize multiple LDR images with different exposures and reconstruct the HDR image using the conventional multi-image technique~\cite{debevec1997recovering}.
However, predicting LDR images with different exposures from a single LDR input itself is challenging as it involves the non-linear CRF mapping, dequantization, and hallucination.

Unlike~\cite{endo2017drtmo, lee2018deep}, our method directly reconstructs an HDR image by modeling the inverse process of the image formation pipeline. 
\figref{pipeline} illustrates the LDR image formation pipeline, state-of-the-art single-image HDR reconstruction approaches~\cite{eilertsen2017hdrcnn, endo2017drtmo, marnerides2018expandnet}, and the proposed method.

\heading{Dequantization and decontouring.}
When converting real-valued HDR images to 8-bit LDR images, quantization errors inevitably occurs. 
They often cause scattered noise or introduce false edges (known as contouring or banding artifacts) particularly in regions with smooth gradient changes.
While these errors may not be visible in the non-linear LDR image, the tone mapping operation (for visualizing an HDR image) often aggravates them, resulting in noticeable artifacts.
Existing decontouring methods smooth images by applying the adaptive spatial filter~\cite{daly2004decontouring} or selective average filter~\cite{song2016hardware}. 
However, these methods often involve meticulously tuned parameters and often produce undesirable artifacts in textured regions.
CNN-based methods have also been proposed~\cite{hou2017image,liu2018learning,zhao2019deep}.
Their focus is on restoring an 8-bit image from lower bit-depth input (e.g., 2-bit or 4-bit).
In contrast, we aim at recovering a 32-bit floating-point image from an 8-bit LDR input image.

\heading{Radiometric calibration.}
As the goal of HDR reconstruction is to measure the full scene irradiance from an input LDR image, it is necessary to estimate the CRF.
Recovering the CRF from a single image requires certain assumptions of statistical priors, e.g., color mixtures at edges~\cite{lin2004radiometric, lin2005determining, ng2007using} or noise distribution~\cite{matsushita2007radiometric, takamatsu2008estimating}.
Nevertheless, these priors may not be applicable to a wide variety of images in the wild.
A CRF can be empirically modeled by the basis vectors extracted from a set of real-world CRFs~\cite{grossberg2003space} via the principal component analysis (PCA).
Li and Peers~\cite{li2017crfnet} train a CRF-Net to estimate the weights of the basis vectors from a single input image and then use the principal components to reconstruct the CRF. 
Our work improves upon~\cite{li2017crfnet} by introducing new features and monotonically increasing constraint.
We show that an accurate CRF is crucial to the quality of the reconstructed HDR image.
After obtaining an accurate HDR image, users can adopt advanced tone-mapping methods (e.g.,~\cite{liang2018hybrid,photomatix}) to render a more visually pleasing LDR image.
Several other applications (e.g., image-based lighting~\cite{debevec2006image} and motion blur synthesis~\cite{debevec1997recovering}) also require linear HDR images for further editing or mapping.

\heading{Image completion.}
Recovering the missing contents in saturated regions can be posed as an image completion problem.
Early image completion approaches synthesize the missing contents via patch-based synthesis~\cite{barnes2009patchmatch,efros2001image,huang2014image}.
Recently, several learning-based methods have been proposed to synthesize the missing pixels using CNNs~\cite{iizuka2017globally,liu2018image,pathak2016context,yu2018free,yu2018generative}. 
Different from the generic image completion task, the missing pixels in the over-exposed regions always have equal or larger values than other pixels in an image.
We incorporate this constraint in our Hallucination-Net to reflect the physical formation in over-exposed regions.

\heading{Camera pipeline.}
We follow the forward LDR image formation pipeline in HDR reconstruction~\cite{debevec1997recovering} and radiometric calibration~\cite{chakrabarti2014modeling} algorithms.
While the HDRCNN method~\cite{eilertsen2017hdrcnn} also models a similar LDR image formation, this model does not learn to estimate accurate CRFs and reduce quantization artifacts.
There exist more advanced and complex camera pipelines to model the demosaicing, white balancing, gamut mapping, noise reduction steps for image formation~\cite{brown2015understanding,karaimer2016software,kim2012new}. 
In this work, we focus on the components of great importance for HDR image reconstruction and model the rest of the pipeline by a refinement network.

\section{Learning to Reverse the Camera Pipeline}
\label{sec:algorithm}

\begin{figure}[t]
    \centering
    \footnotesize
    \includegraphics[width=1.0\linewidth]{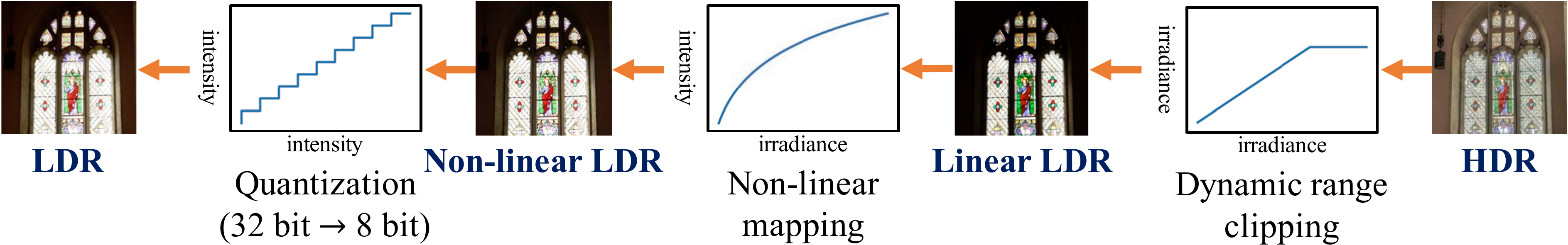}
    \\
    (a) LDR Image formation pipeline
    \\
    \vspace{1mm}
    \includegraphics[width=1.0\linewidth]{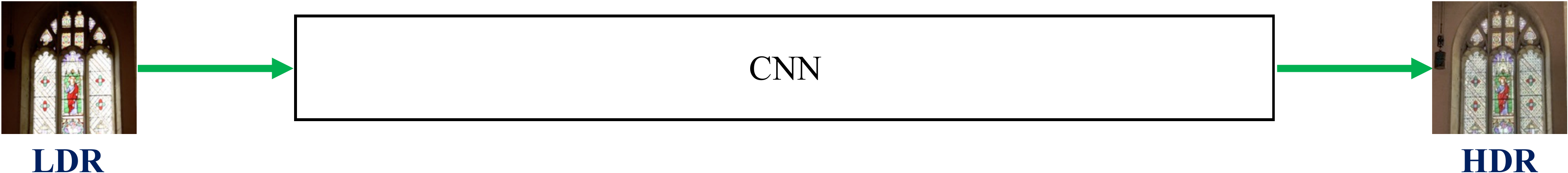}\\
    (b) ExpandNet~\cite{marnerides2018expandnet}
    \\
    \vspace{1mm}
    \includegraphics[width=1.0\linewidth]{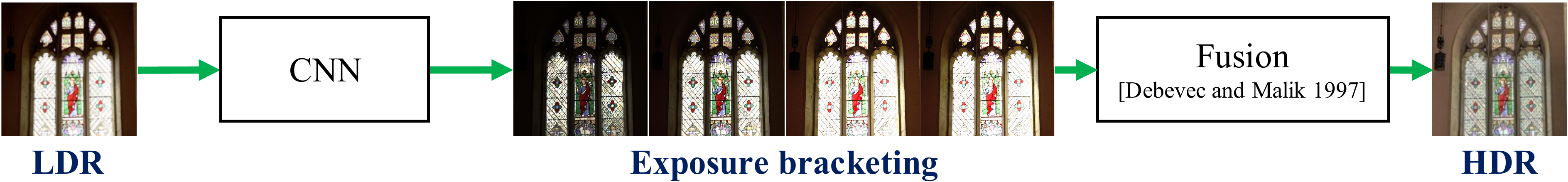}\\
    (c) DrTMO~\cite{endo2017drtmo}
    \\
    \vspace{1mm}
    \includegraphics[width=1.0\linewidth]{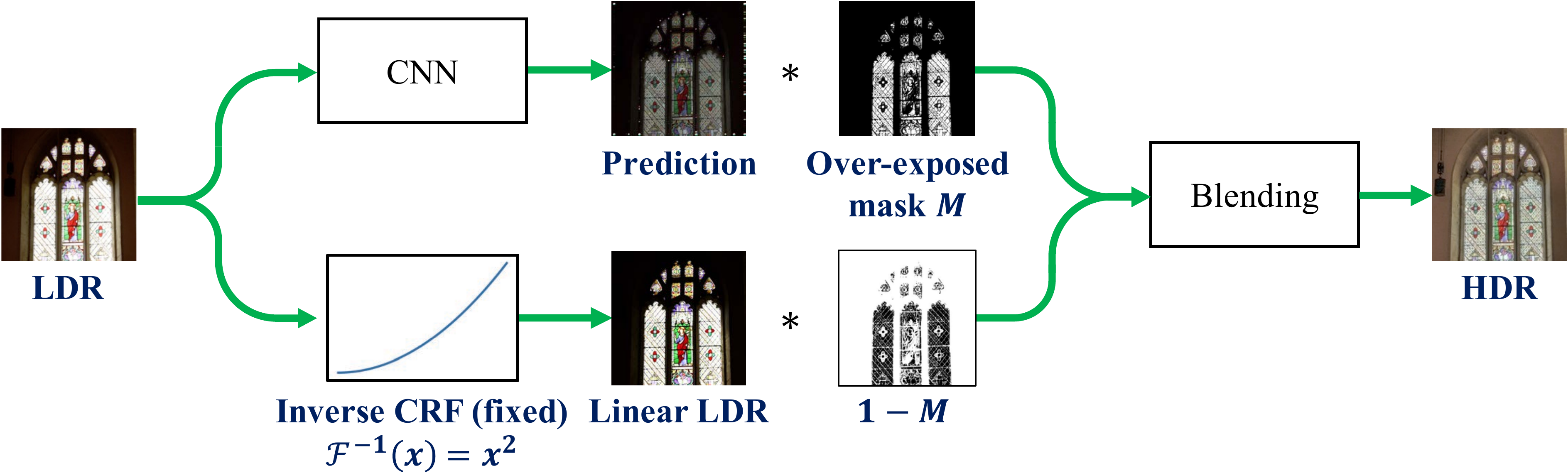}\\
    (d) HDRCNN~\cite{eilertsen2017hdrcnn}
    \\
    \vspace{1mm}
    \includegraphics[width=1.0\linewidth]{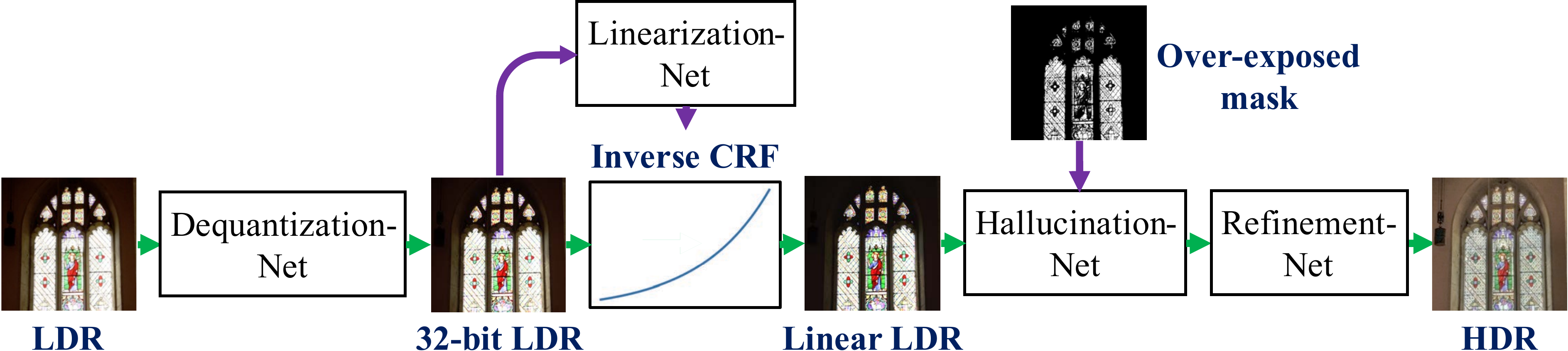}\\
    (e) Proposed method
    \vspace{-1mm}
    \caption{
    \textbf{The LDR Image formation pipeline and overview of single-image HDR reconstruction methods.} 
    (a) We model the LDR image formation by (from right to left) dynamic range clipping, non-linear mapping, and quantization~\cite{debevec1997recovering}. 
    (b) ExpandNet~\cite{marnerides2018expandnet} learns a direct mapping from LDR to HDR images.
    (c) DrTMO~\cite{endo2017drtmo} synthesizes multiple LDR images with different exposures and fuses them into an HDR image.
    (d) HDRCNN~\cite{eilertsen2017hdrcnn} predicts details in over-exposed regions while ignoring the quantization errors in the under-exposed regions.
    (e) The proposed method explicitly learns to ``reverse" each step of the LDR image formation pipeline.
    }
  \label{fig:pipeline}
  \figmargin
\end{figure}

In this section, we first introduce the image formation pipeline that renders an LDR image from an HDR image (the scene irradiance) as shown in \figref{pipeline}(a).
We then describe our design methodology and training procedures for single-image HDR reconstruction by reversing the image formation pipeline as shown in \figref{pipeline}(e).

\subsection{LDR image formation}
\label{sec:formation}
While the real scene irradiance has a high dynamic range, the digital sensor in cameras can only capture and store a limited extent, usually with 8 bits.
Given the irradiance $\irradiance$ and sensor exposure time $\exposure$, an HDR image is recorded by $\hdr = \irradiance \times \exposure$.
The process of converting one HDR image to one LDR image can be modeled by the following major steps:

\noindent{(1) \textbf{Dynamic range clipping}.}
The camera first clips the pixel values of an HDR image $\hdr$ to a limited range, which can be formulated by $\cliphdr = \clipping{\hdr} = \min{(\hdr, 1)}$.
Due to the clipping operation, there is information loss for pixels in the over-exposed regions. 

\noindent{(2) \textbf{Non-linear mapping}.}
To match the human perception of the scene, a camera typically applies a non-linear CRF mapping to adjust the contrast of the captured image: $\nonlinearhdr =  \crfmapping{\cliphdr}$.
A CRF is unique to the camera model and unknown in our problem setting.

\noindent{(3) \textbf{Quantization}.}
After the non-linear mapping, the recorded pixel values are quantized to 8 bits by 
$\quantizing{\nonlinearhdr} = \left \lfloor 255 \times \nonlinearhdr + 0.5 \right \rfloor / 255$. 
The quantization process leads to errors in the under-exposed and smooth gradient regions.

In summary, an LDR image $\ldr$ is formed by:
\begin{equation} \label{eq:pipeline}
    \ldr = \applypipeline{\hdr} = \quantizing{\crfmapping{\clipping{\hdr}}}\,,
\end{equation}
where $\pipeline$ denotes the pipeline of dynamic range clipping, non-linear mapping, and quantization steps.

To learn the inverse mapping $\invpipeline$, we propose to decompose the HDR reconstruction task into three sub-tasks: dequantization, linearization, and hallucination, which model the inverse functions of the quantization, non-linear mapping, and dynamic range clipping, respectively.
We train three CNNs for the three sub-tasks using the corresponding supervisory signal and specific physical constraints.
We then integrate these three networks into an end-to-end model and jointly fine-tune to further reduce error accumulation and improve the performance.

\subsection{Dequantization}
Quantization often results in scattered noise or contouring artifacts in smooth regions.
Therefore, we propose to learn a Dequantization-Net to reduce the quantization artifacts in the input LDR image.

\heading{Architecture.}
Our Dequantization-Net adopts a 6-level U-Net architecture.
Each level consists of two convolutional layers followed by a leaky ReLU ($\alpha = 0.1$) layer.
We use the \texttt{Tanh} layer to normalize the output of the last layer to $[-1.0, 1.0]$.
Finally, we add the output of the Dequantization-Net to the input LDR image to generate the dequantized LDR image $\deqout$.

\heading{Training.}
We minimize the $\ltwo$ loss between the dequantized LDR image $\deqout$ and corresponding ground-truth image $\nonlinearhdr$: $\deqloss = \normtwo{\deqout - \nonlinearhdr}$.
Note that $\nonlinearhdr = \crfmapping{\clipping{\hdr}}$ is constructed from the ground-truth HDR image with dynamic range clipping and non-linear mapping.

\subsection{Linearization}
The goal of linearization (i.e., radiometric calibration) is to estimate a CRF and convert a non-linear LDR image to a linear irradiance.
Although the CRF (denoted by $\crf$) is distinct for each camera, all the CRFs must have the following properties.
First, the function should be monotonically increasing.
Second, the minimal and maximal input values should be respectively mapped to the minimal and maximal output values: $\crfmapping{0} = 0$ and $\crfmapping{1} = 1$ in our case.
As the CRF is a one-to-one mapping function, the inverse CRF (denoted by $\invcrf = \crf^{-1}$) also has the above properties.

To represent a CRF, we first discretize an inverse CRF by uniformly sampling 1024 points between $[0, 1]$.
Therefore, an inverse CRF is represented as a 1024-dimensional vector $\invcrfvector \in \mathbb{R}^{1024}$.
We then adopt the Empirical Model of Response (EMoR) model~\cite{grossberg2003space}, which assumes that each inverse CRF $\invcrfvector$ can be approximated by a linear combination of $K$ PCA basis vectors.
In this work, we set $K = 11$ as it has been shown to capture the variations well in the CRF dataset~\cite{li2017crfnet}.
To predict the inverse CRF, we train a Linearization-Net to estimate the weights from the input non-linear LDR image.

\begin{figure}
    \centering
    \includegraphics[width=1\columnwidth]{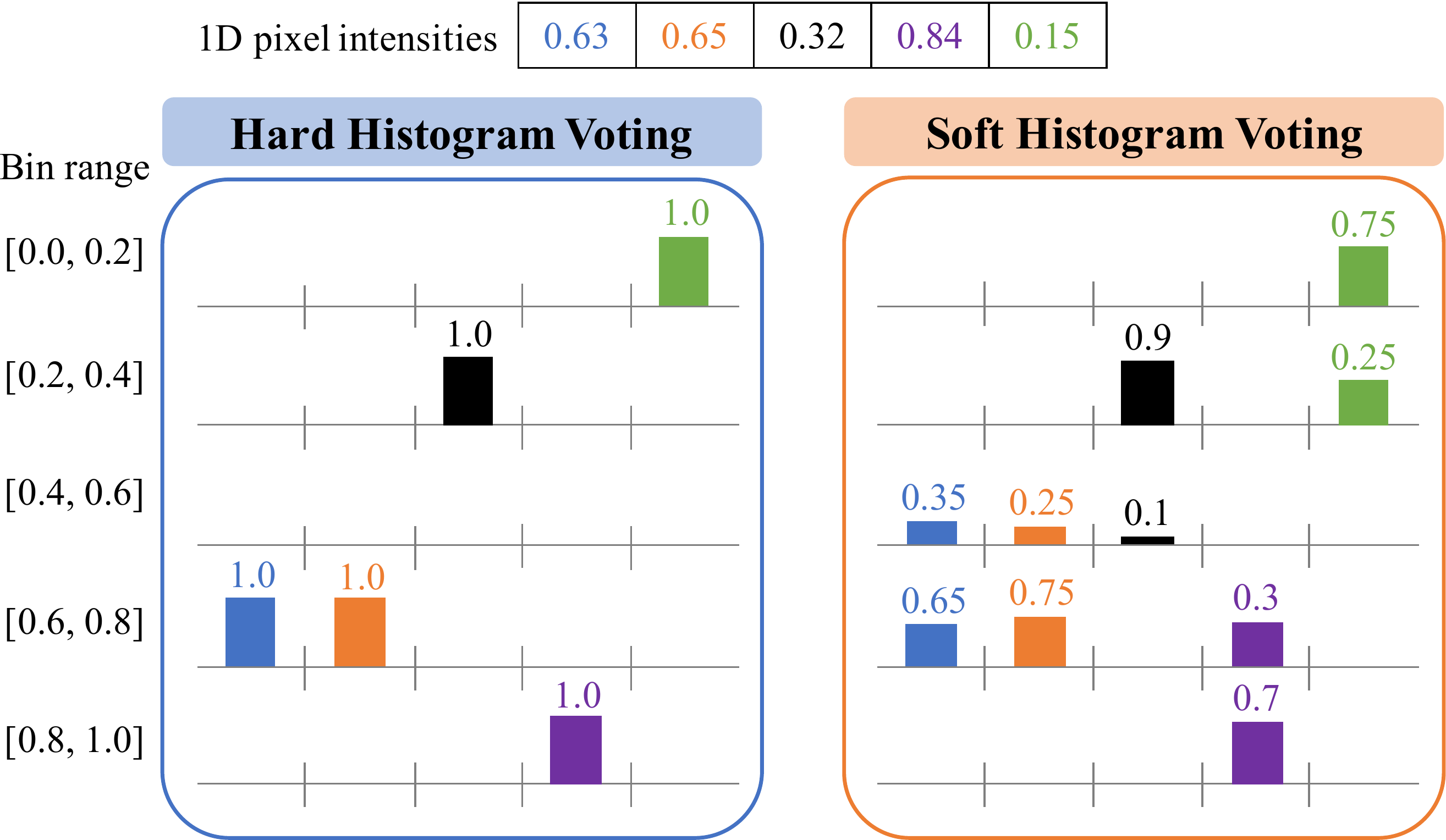}
    \figmargin
    \caption{
        \textbf{Spatial-aware soft histogram layer.}
        We extract histogram features by soft counting on pixel intensities and preserving the spatial information.
    }
    \label{fig:histogram_feature}
    \figmargin
\end{figure}

\heading{Input features.}
As the edge and color histogram have been shown effective to estimate an inverse CRF~\cite{lin2004radiometric,lin2005determining}, we first extract the edge and histogram features from the non-linear LDR image.
We adopt the Sobel filter to obtain the edge responses, resulting in 6 feature maps (two directions $\times$ three color channels).
To extract the histogram features, we propose a \emph{spatial-aware soft-histogram layer}.
Specifically, given the number of histogram bins $B$, we construct a \emph{soft} counting of pixel intensities by:
\begin{align} \label{eq:histogram_layer}
    h(i,j,c,b) = 
    \begin{cases}
        1 - d \cdot B\,, & \text{if}~ d < \frac{1}{B} \\
        0\,, & \text{otherwise}
    \end{cases}
\end{align}
where $i, j$ indicate horizontal and vertical pixel positions, $c$ denotes the index of color channels, $b \in \{1,\cdots, B\}$ is the index for the histogram bin, and $d = | I(i,j,c) - (2b - 1) / (2B) |$ is the intensity distance to the center of the $b$-th bin.
Every pixel contributes to the two nearby bins according to the intensity distance to the center of each bin.
\figref{histogram_feature} shows a 1D example of our soft-histogram layer.
Our histogram layer preserves the spatial information and is fully differentiable.

\begin{figure}
    \centering
    \includegraphics[width=1.0\linewidth]{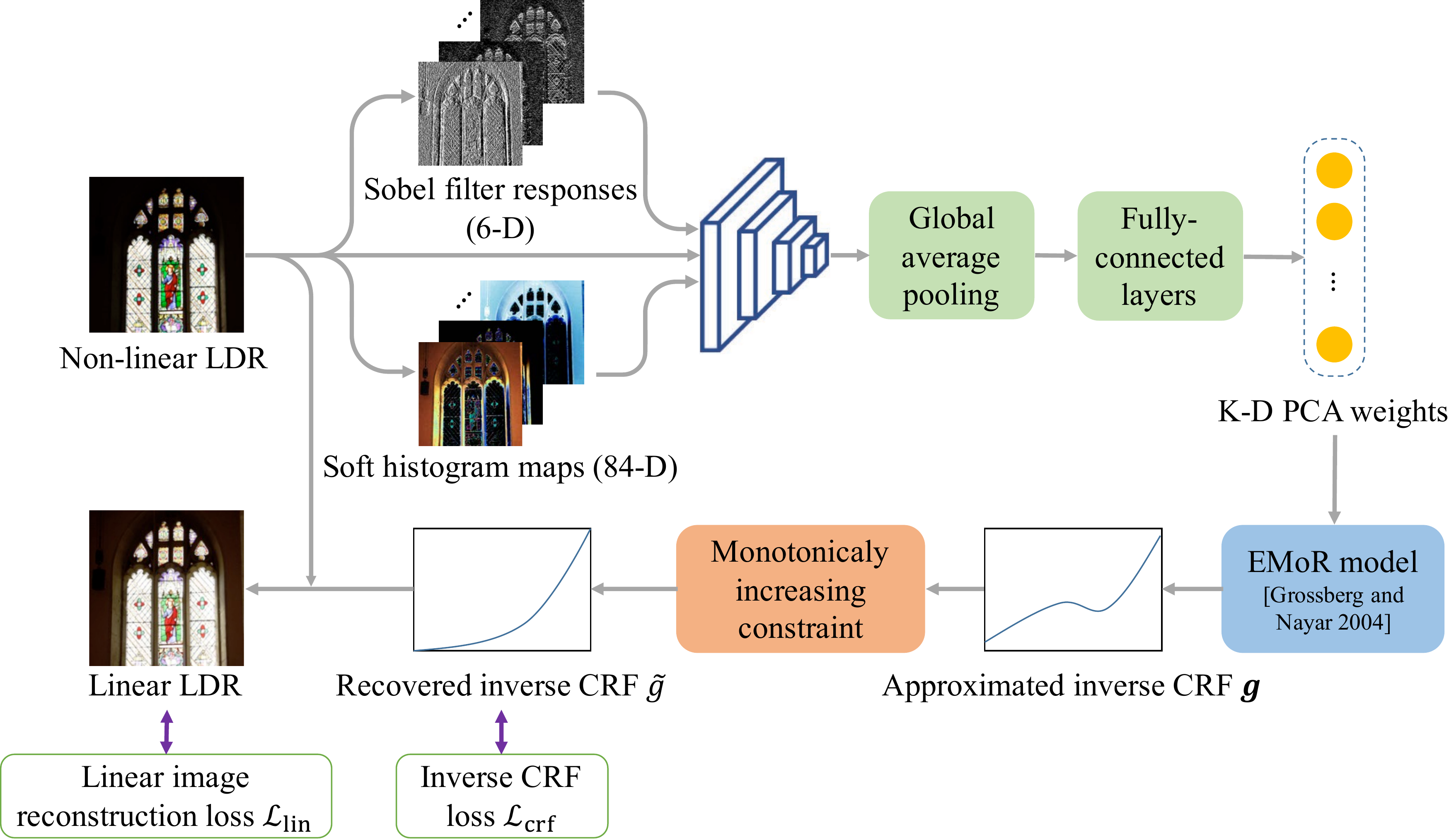}
    \figmargin
    \caption{
        \textbf{Architecture of the Linearization-Net.}
        Our Linearization-Net takes as input the non-linear LDR image, edge maps, and histogram maps, and predicts the PCA coefficients for reconstructing an inverse CRF, followed by enforcing the monotonically increasing constraint.
    }
    \label{fig:lin_net}
    \figmargin
\end{figure}

\heading{Architecture.}
We use the ResNet-18~\cite{he2016deep} as the backbone of our Linearization-Net.
To extract a global feature, we add a global average pooling layer after the last convolutional layer.
We then use two fully-connected layers to generate $K$ PCA weights and reconstruct an inverse CRF.

\heading{Monotonically increasing constraint.}
To satisfy the constraint that a CRF/inverse CRF should be monotonically increasing, we adjust the estimated inverse CRF by enforcing all the first-order derivatives to be non-negative.
Specifically, we calculate the first-order derivatives by $\invcrfder_1 = 0$ and $\invcrfder_d = \invcrfscalar_d - \invcrfscalar_{d-1}$ for $d \in [2, \cdots, 1024]$ and find the smallest negative derivative $\invcrfder_m = \min(\min_d(\invcrfder_d), 0)$.
We then shift the derivatives by $\norminvcrfder_d = \invcrfder_d - \invcrfder_m$.
The inverse CRF $\norminvcrfvector = [\norminvcrfscalar_1, \cdots,  \norminvcrfscalar_{1024}]$ is then reconstructed by integration and normalization:
\begin{align} \label{eq:crf_normalization}
    \norminvcrfscalar_d = \frac{1}{ \sum_{i=1}^{1024} \norminvcrfder_i } \sum_{i=1}^{d} \norminvcrfder_i \,.
\end{align}
We normalize $\norminvcrfscalar_d$ to ensure the inverse CRF satisfies the constraint that $\invcrfmapping{0} = 0$ and $\invcrfmapping{1} = 1$.
\figref{lin_net} depicts the pipeline of our Linearization-Net. 
With the normalized inverse CRF $\norminvcrfvector$, we then map the non-linear LDR image $\deqout$ to a linear LDR image $\linout$.

\heading{Training.}
We define the linear LDR image reconstruction loss by: $\linloss = \normtwo{ \linout - \cliphdr}$,
where $\cliphdr = \clipping{\hdr}$ is constructed from the ground-truth HDR image with the dynamic range clipping process.
In addition, we formulate the inverse CRF reconstruction loss by: $\crfloss = \normtwo{\norminvcrfvector - \invcrfvector}$, 
where $\invcrfvector$ is the ground-truth inverse CRF.
We train the Linearization-Net by optimizing $\linloss + \crflossweight \crfloss$.
We empirically set $\crflossweight = 0.1$ in all our experiments.

\subsection{Hallucination}
After dequantization and linearization, we aim to recover the missing contents due to dynamic range clipping.
To this end, we train a Hallucination-Net (denoted by $\halnet{\cdot}$) to predict the missing details within the over-exposed regions.

\begin{figure}
	\centering
	\includegraphics[width=1.0\linewidth]{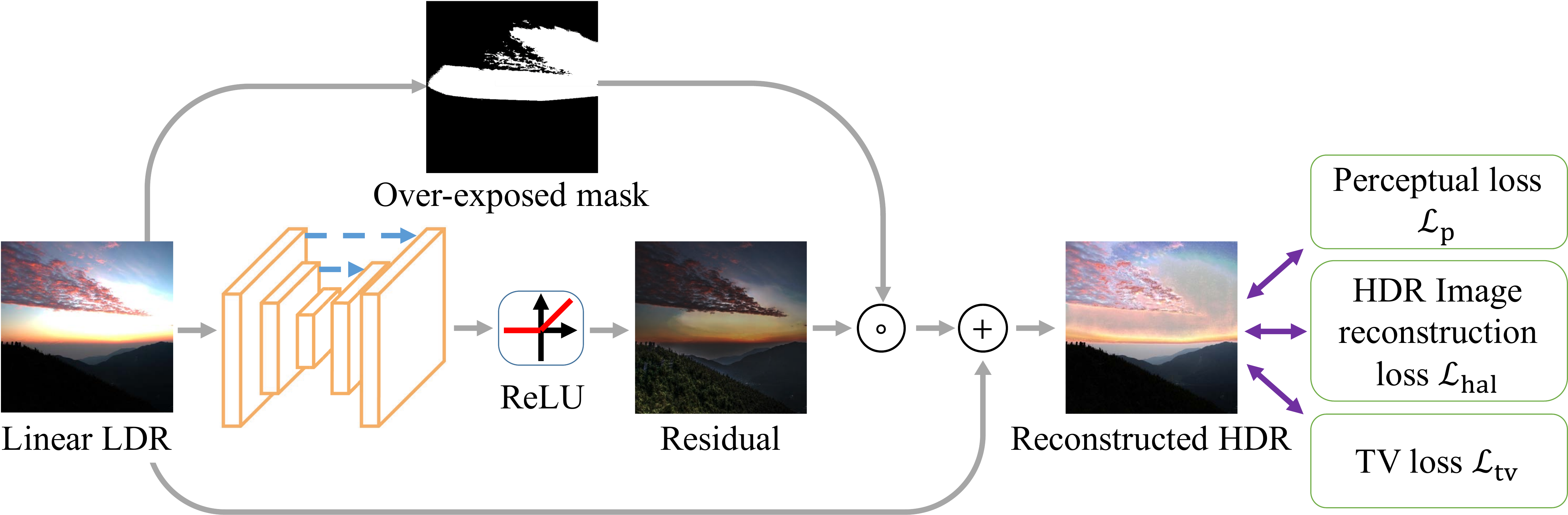}
	\figmargin
	\caption{
		\textbf{Architecture of the Hallucination-Net.}
		We train the Hallucination-Net to predict \emph{positive} residuals and recover missing content in the over-exposed regions.
	}
	\label{fig:hallucination-net}
	\figmargin
\end{figure}

\heading{Architecture.}
We adopt an encoder-decoder architecture with skip connections~\cite{eilertsen2017hdrcnn} as our Hallucination-Net.
The reconstructed HDR image is modeled by $\halout = \linout + \mask \cdot \halnet{\linout}$, 
where $\linout$ is the image generated from the Linearization-Net and $\mask = \max(0, \linout - \gamma) / (1 - \gamma)$ is the over-exposed mask with $\gamma = 0.95$.
Since the missing values in the over-exposed regions should always be greater than the existing pixel values, we constrain the Hallucination-Net to predict \emph{positive residuals} by adding a \texttt{ReLU} layer at the end of the network.
We note that our over-exposed mask is a \emph{soft} mask where $\mask \in [0, 1]$.
The soft mask allows the network to smoothly blend the residuals with the existing pixel values around the over-exposed regions.
\figref{hallucination-net} shows the design of our Hallucination-Net.

We find that the architecture of~\cite{eilertsen2017hdrcnn} may generate visible checkerboard artifacts in large over-exposed regions.
In light of this, we replace the transposed convolutional layers in the decoder with the resize-convolution layers~\cite{odena2016deconvolution}.

\heading{Training.}
We train our Hallucination-Net by minimizing the $\logltwo$ loss: $\halloss = \normtwo{\log(\halout) - \log(\hdr)}$,
where $\hdr$ is the ground-truth HDR image.
We empirically find that training is more stable and achieves better performance when minimizing the loss in the log domain.
As the highlight regions (e.g., sun and light sources) in an HDR image typically have values with orders of magnitude larger than those of other regions, the loss is easily dominated by the errors in the highlight regions when measured in the linear domain.
Computing the loss in the log domain reduces the influence of these extremely large errors and encourages the network to restore more details in other regions.

To generate more realistic details, we further include the perceptual loss $\perceptualloss$~\cite{johnson2016perceptual}:
As the VGG-Net (used in $\perceptualloss$) is trained on \emph{non-linear RGB images}, directly feeding an linear HDR image to the VGG-Net may not obtain meaningful features.
Therefore, we first apply a differentiable global tone-mapping operator~\cite{wu2018deep} to map the HDR images to a non-linear RGB space.
We can then compute the perceptual loss on the tone-mapped HDR images.
To improve the spatial smoothness of the predicted contents, we also minimize the total variation (TV) loss $\tvloss$ on the recovered HDR image.
Our total loss for training the Hallucination-Net is $\halloss + \perceptuallossweight \perceptualloss + \tvlossweight \tvloss$.
We empirically set $\perceptuallossweight = 0.001$ and $\tvlossweight = 0.1$ in our experiments.

\subsection{Joint training}
\label{sec:joint_training}

We first train the Dequantization-Net, Linearization-Net, and Hallucination-Net with the corresponding input and ground-truth data.
After the three networks converge, we jointly fine-tune the entire pipeline by minimizing the combination of loss functions $\totalloss$:
\begin{align} \label{eq:total_loss}
	 \deqlossweight \deqloss\!+\!\linlossweight \linloss\!+\!\crflossweight \crfloss\!+\!\hallossweight \halloss\!+\!\perceptuallossweight \perceptualloss\!+\!\tvlossweight \tvloss\,
\end{align}
where we set the weights to $\deqlossweight = 1$, $\linlossweight = 10$, $\crflossweight = 1$, $\hallossweight = 1$, $\perceptuallossweight = 0.001$, and $\tvlossweight = 0.1$.
The joint training reduces error accumulation between the sub-networks and further improves the reconstruction performance.

\subsection{Refinement}
\label{sec:refinement}
Modern camera pipeline contains significant amounts of \emph{spatially-varying} operations, e.g. local tone-mapping, sharpening, chroma denoising, lens shading correction, and white balancing.
To handle these effects that are not captured by our image  formation pipeline, we introduce an optional Refinement-Net.

\heading{Architecture.}
Our Refinement-Net adopts the same U-Net architecture as the Dequantization-Net, which learns to refine the output of the Hallucination-Net by a residual learning.
The output of the Refinement-net is denoted by $\reout$.

\heading{Training.}
To model the effects of real camera pipelines, we train the Refinement-Net using HDR images reconstructed from exposure stacks captured by various cameras (more details in the supplementary material).
We minimize the same $\totalloss$ for end-to-end fine-tuning (with $\deqlossweight$, $\linlossweight$, $\crflossweight$, and $\hallossweight$ set to 0 as there are no stage-wise supervisions), and replace the output of Hallucination-Net $\halout$ with refined HDR image $\reout$.

\section{Experimental Results}
\label{sec:experiments}

We first describe our experimental settings and evaluation metrics.
Next, we present quantitative and qualitative comparisons with the state-of-the-art single-image HDR reconstruction algorithms.
We then analyze the contributions of individual modules to justify our design choices.

\subsection{Experiment setups}
\label{sec:exp_setting}

\heading{Datasets.} 
For training and evaluating single-image HDR reconstruction algorithms, we construct two HDR image datasets: \textsc{HDR-Synth} and \textsc{HDR-Real}.
We also evaluate our method on two publicly available datasets: \textsc{RAISE} (RAW-jpeg pairs)~\cite{dang2015raise} and \textsc{HDR-Eye}~\cite{nemoto2015visual}.

\heading{Evaluation metrics.}
We adopt the HDR-VDP-2~\cite{mantiuk2011hdr} to evaluate the accuracy of HDR reconstruction. 
We normalize both the predicted HDR and reference ground-truth HDR images with the processing steps in~\cite{marnerides2018expandnet}.
We also evaluate the PSNR, SSIM, and perceptual score with the LPIPS metric~\cite{zhang2018perceptual} on the tone-mapped HDR images in the supplementary material.

\begin{table*}[t]
    \caption{
        \textbf{Quantitative comparison on HDR images with existing methods.} %
        * represents that the model is re-trained on our synthetic training data and + is fine-tuned on both synthetic and real training data. \textcolor{red}{\pmb{Red}} text indicates the best and \textcolor{blue}{\underline{blue}} text indicates the best performing state-of-the-art method.
    }
    \label{tab:compare_stoa_hdrvdp}
    \centering
    \footnotesize
    \renewcommand{\tabcolsep}{4pt} %
    \vspace{1mm}
    \begin{tabular}{l|l|cccc}
        \toprule
        Method & Training dataset & \textsc{HDR-Synth} & \textsc{HDR-Real} & \textsc{RAISE}~\cite{dang2015raise} & \textsc{HDR-Eye}~\cite{nemoto2015visual} \\
        \midrule
        HDRCNN+~\cite{eilertsen2017hdrcnn} & \textsc{HDR-Synth} + \textsc{HDR-Real} & $55.51\pm6.64$ & \textcolor{blue}{\underline{$51.38\pm7.17$}} & $56.51\pm4.33$ & $51.08\pm5.84$ \\
        DrTMO+~\cite{endo2017drtmo} & \textsc{HDR-Synth} + \textsc{HDR-Real} & \textcolor{blue}{\underline{$56.41\pm7.20$}} & $50.77\pm7.78$ & \textcolor{blue}{\underline{$57.92\pm3.69$}} & \textcolor{blue}{\underline{$51.26\pm5.94$}} \\
        ExpandNet~\cite{marnerides2018expandnet} & Pre-trained model of~\cite{marnerides2018expandnet} & $53.55\pm4.98$ & $48.67\pm6.46$ & $54.62\pm1.99$ & $50.43\pm5.49$ \\
        Deep chain HDRI~\cite{lee2018deepchain} & Pre-trained model of~\cite{lee2018deepchain} & - & - & - & $49.80\pm5.97$ \\
        Deep recursive HDRI~\cite{lee2018deep} & Pre-trained model of~\cite{lee2018deep} & - & - & - & $48.85\pm4.91$ \\
        Ours* & \textsc{HDR-Synth} & \textcolor{red}{$\pmb{60.11\pm6.10}$} & $51.59\pm7.42$ & $58.80\pm3.91$ & $52.66\pm5.64$\\
        Ours+ & \textsc{HDR-Synth} + \textsc{HDR-Real} & $59.52\pm6.02$ & \textcolor{red}{\pmb{$53.16\pm7.19$}} & \textcolor{red}{\pmb{$59.21\pm3.68$}} & \textcolor{red}{\pmb{$53.16\pm5.92$}} \\
        \bottomrule
    \end{tabular}
    \vspace{2mm}
\end{table*}

\begin{figure*}
	\centering
	\footnotesize
    \renewcommand{\tabcolsep}{1pt} %
	\renewcommand{\arraystretch}{1} %
	\newcommand{\imagewidth}{0.195\textwidth}
	\newcommand{\patchwidth}{0.095\textwidth}
	\newcommand{\addimages}[1]{
	    \multicolumn{2}{c}{\includegraphics[width=\imagewidth]{imgs/SOTA/#1/A_box.jpg}} & \multicolumn{2}{c}{\includegraphics[width=\imagewidth]{imgs/SOTA/#1/B_box.jpg}} & \multicolumn{2}{c}{\includegraphics[width=\imagewidth]{imgs/SOTA/#1/C_box.jpg}} & \multicolumn{2}{c}{\includegraphics[width=\imagewidth]{imgs/SOTA/#1/D_box.jpg}} & \multicolumn{2}{c}{\includegraphics[width=\imagewidth]{imgs/SOTA/#1/E_box.jpg}}
	    \\
	    \includegraphics[width=\patchwidth]{imgs/SOTA/#1/A_patch1.png} & \includegraphics[width=\patchwidth]{imgs/SOTA/#1/A_patch2.png} & \includegraphics[width=\patchwidth]{imgs/SOTA/#1/B_patch1.png} & \includegraphics[width=\patchwidth]{imgs/SOTA/#1/B_patch2.png} & \includegraphics[width=\patchwidth]{imgs/SOTA/#1/C_patch1.png} & \includegraphics[width=\patchwidth]{imgs/SOTA/#1/C_patch2.png} & \includegraphics[width=\patchwidth]{imgs/SOTA/#1/D_patch1.png} & \includegraphics[width=\patchwidth]{imgs/SOTA/#1/D_patch2.png} & \includegraphics[width=\patchwidth]{imgs/SOTA/#1/E_patch1.png} & \includegraphics[width=\patchwidth]{imgs/SOTA/#1/E_patch2.png}
	}
	\begin{tabular}{cccccccccc}
	    \multicolumn{2}{c}{\includegraphics[width=\imagewidth]{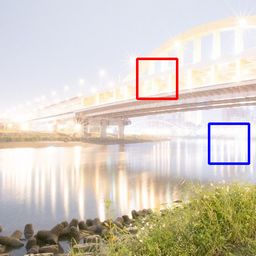}} & \multicolumn{2}{c}{\includegraphics[width=\imagewidth]{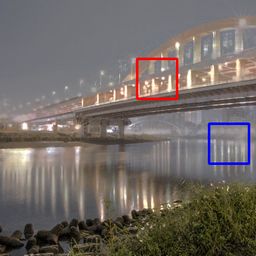}} & \multicolumn{2}{c}{\includegraphics[width=\imagewidth]{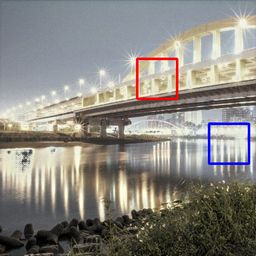}} & \multicolumn{2}{c}{\includegraphics[width=\imagewidth]{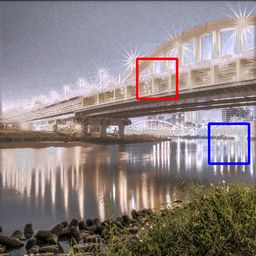}} & \multicolumn{2}{c}{\includegraphics[width=\imagewidth]{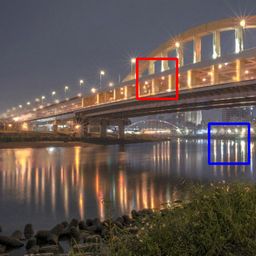}}
	    \\
	    \includegraphics[width=\patchwidth]{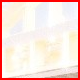} & \includegraphics[width=\patchwidth]{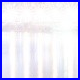} & \includegraphics[width=\patchwidth]{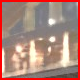} & \includegraphics[width=\patchwidth]{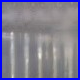} & \includegraphics[width=\patchwidth]{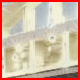} & \includegraphics[width=\patchwidth]{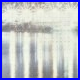} & \includegraphics[width=\patchwidth]{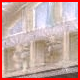} & \includegraphics[width=\patchwidth]{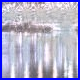} & \includegraphics[width=\patchwidth]{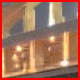} & \includegraphics[width=\patchwidth]{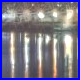} \\
	    \addimages{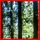} \\
	    \multicolumn{2}{c}{(a) Input LDR} & 
	    \multicolumn{2}{c}{(b) HDRCNN+} & 
	    \multicolumn{2}{c}{(c) DrTMO+} & 
	    \multicolumn{2}{c}{(d) ExpandNet} & 
	    \multicolumn{2}{c}{(e) Ours+}
	\end{tabular}
	\caption{
		\textbf{Visual comparison on real input image.}
		The example on the top is captured by NIKON D90 from \textsc{HDR-Real}, and the bottom one is from DrTMO~\cite{endo2017drtmo}.
		The HDRCNN~\cite{eilertsen2017hdrcnn} often suffers from noise, banding artifacts or over-saturated colors in the under-exposed regions.
		The DrTMO~\cite{endo2017drtmo} cannot handle over-exposed regions well and leads to blurry and low-contrast results.
		The ExpandNet~\cite{marnerides2018expandnet} generates artifacts in the over-exposed regions. 
		In contrast, our method restores fine details in both the under-exposed and over-exposed regions and renders visually pleasing results.
	} 
	\label{fig:visual_sota}
	\vspace{2mm}
\end{figure*}

\subsection{Comparisons with state-of-the-art methods}
\label{sec:compare_SOTA}

We compare the proposed method with five recent CNN-based approaches: HDRCNN~\cite{eilertsen2017hdrcnn}, DrTMO~\cite{endo2017drtmo}, ExpandNet~\cite{marnerides2018expandnet}, Deep chain HDRI~\cite{lee2018deepchain}, and Deep recursive HDRI~\cite{lee2018deep}. 
As the ExpandNet does not provide the code for training, we only compare with their released pre-trained model.
Both the Deep chain HDRI and Deep recursive HDRI methods do not provide their pre-trained models, so we compare with the results on the \textsc{HDR-Eye} dataset reported in their papers.

We first train our model on the training set of the \textsc{HDR-Synth} dataset (denoted by Ours) and the fine-tune on the training set of the \textsc{HDR-Real} dataset (denoted by Ours+).
For fair comparisons, we also re-train the HDRCNN and DrTMO models with both the \textsc{HDR-Synth} and \textsc{HDR-Real} datasets (denoted by HDRCNN+ and DrTMO+).
We provide more comparisons with the pre-trained models of HDRCNN and DrTMO and the our results from each training stage in the supplementary material.

\heading{Quantitative comparisons.}
\tabref{compare_stoa_hdrvdp} shows the average HDR-VDP-2 scores on the \textsc{HDR-Synth}, \textsc{HDR-Real}, \textsc{RAISE}, and \textsc{HDR-Eye} datasets.
The proposed method performs favorably against the state-of-the-art methods on all four datasets.
After fine-tuning on the \textsc{HDR-Real} training set, the performance of our model (Ours+) is further improved by 1.57 on \textsc{HDR-Real}, 0.41 on the \textsc{RAISE}, and 0.5 on \textsc{HDR-Eye} datasets, respectively.

\heading{Visual comparisons.} 
\figref{visual_sota} compares the proposed model with existing methods on a real image captured using NIKON D90 provided by \textsc{HDR-Real} and an example provided in~\cite{endo2017drtmo}.
We note that both two examples in~\figref{visual_sota} come from unknown camera pipeline, and there are no ground-truth HDRs.
In general, the HDRCNN~\cite{eilertsen2017hdrcnn} often generates overly-bright results and suffers from noise in the under-exposed regions as an aggressive and fixed inverse CRF $x^2$ is used. 
The results of the DrTMO~\cite{endo2017drtmo} often looks blurry or washed-out.
The ExpandNet~\cite{marnerides2018expandnet} cannot restore the details well in the under-exposed regions and generates visual artifacts in the over-exposed regions, such as sky.
Due to the space limit, we provide more visual comparisons in the supplementary material.

\heading{User study.}
We conduct a user study to evaluate the human preference on HDR images. 
We adopt the paired comparison~\cite{Lai-CVPR-2016, Rubinstein-TOG-2010}, where users are asked to select a preferred image from a pair of images in each comparison.
We design the user study with the following two settings:
(1) \textit{With-reference test}: We show both the input LDR and the ground-truth HDR images as reference.
This test evaluates the \emph{faithfulness} of the reconstructed HDR image to the ground-truth.
(2) \textit{No-reference test}: The input LDR and ground-truth HDR images are not provided. %
This test mainly compares the \emph{visual quality} of two reconstructed HDR images.

\begin{figure}[t]
\centering
\includegraphics[width=0.98\columnwidth]{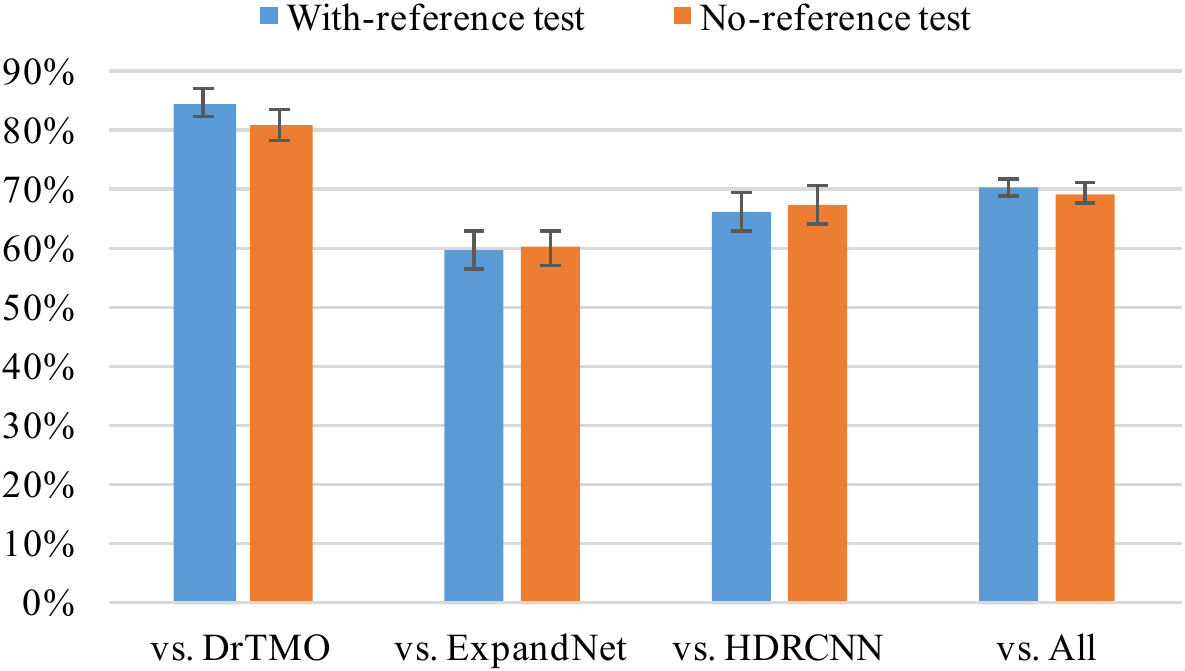}
    \figmargin
\caption{
    \textbf{Results of user study.} Our results are preferred by users in both with-reference and no-reference tests.
    }
\label{fig:user_study}
    \vspace{0mm}
\end{figure}

We evaluate all 70 HDR images in the \textsc{HDR-Real} test set. 
We compare the proposed method with the HDRCNN~\cite{eilertsen2017hdrcnn}, DrTMO~\cite{endo2017drtmo}, and ExpandNet~\cite{marnerides2018expandnet}.
We ask each participant to compare 30 pairs of images and collect the results from a total of 200 unique participants.
\figref{user_study} reports the percentages of the head-to-head comparisons in which users prefer our method over the HDRCNN, DrTMO, and ExpandNet.
Overall, there are $70\%$ and $69\%$ of users prefer our results in the with-reference and no-reference tests, respectively.
Both user studies show that the proposed method performs well to human subjective perception.

\begin{table}
    \centering
    \footnotesize
    \caption{
        \textbf{Comparisons on Dequantization-Net.}
        Our Dequantization-Net restores the missing details due to quantization and outperforms existing methods.
    }
    \label{tab:compare_deq}
    \begin{tabular}{l|cc}
        \toprule
        Method & PSNR ($\uparrow$) & SSIM ($\uparrow$) \\
        \midrule
        w/o dequantization 			& $33.86\pm6.96$ & $0.9946\pm0.0109$ \\
        Hou et al.~\cite{hou2017image} 		& $33.79\pm6.72$ & $0.9936\pm0.0110$ \\
        Liu et al.~\cite{liu2018learning} 		& $34.83\pm6.04 $ & $0.9954\pm0.0073$ \\
        Dequantization-Net (Ours) 	& $\pmb{35.87\pm6.11}$ & $\pmb{0.9955\pm0.0070}$  \\ 
        \bottomrule
    \end{tabular}
    \tabmargin
\end{table}

\subsection{Ablation studies}
\label{sec:ablation}

In this section, we evaluate the contributions of individual components using the \textsc{HDR-Synth} test set.

\heading{Dequantization.} 
We consider the LDR images as the input and the image $\nonlinearhdr = \crfmapping{\clipping{\hdr}}$ synthesized from the HDR images as the ground-truth of the dequantization procedure.
We compare our Dequantization-Net with CNN-based models~\cite{hou2017image,liu2018learning}.
\tabref{compare_deq} shows the quantitative comparisons of dequantized images, where our method performs better than other approaches.

\heading{Linearization.} 
Our Linearization-Net takes as input the non-linear LDR image, Sobel filter responses, and histogram features to estimate an inverse CRF.
To validate the effectiveness of these factors, we train our Linearization-Net with different combinations of the edge and histogram features.
\tabref{compare_lin_features} shows the reconstruction error of the inverse CRF and the PSNR between the output of our Linearization-Net and the corresponding ground-truth image $\cliphdr = \clipping{\hdr}$.
The edge and histogram features help predict more accurate inverse CRFs.
The monotonically increasing constraint further boosts the reconstruction performance on both the inverse CRFs and the linear images.

\begin{table}[t]
    \centering
    \caption{
        \textbf{Analysis on alternatives of Linearization-Net.}
        We demonstrate the edge and histogram features and monotonically increasing constraint are effective to improve the performance of our Linearization-Net.
    }
    \label{tab:compare_lin_features}
    \resizebox{\linewidth}{!}{
    \begin{tabular}{cccc|cc}
        \toprule
        \multirow{2}{*}{Image} & \multirow{2}{*}{Edge} & \multirow{2}{*}{Histogram} & Monotonically & L2 error ($\downarrow$) & PSNR ($\uparrow$)
        \\
        & & & increasing & of inverse CRF & of linear image 
        \\
        \midrule
        \checkmark & - & - & - & $2.00\pm3.15$ & $33.43\pm7.03$ \\
        \checkmark & \checkmark & - & - & $1.66\pm2.93$ & $34.31\pm6.94$ \\
        \checkmark & - & \checkmark & - & $1.61\pm3.03$ & $34.51\pm7.14$ \\
        \checkmark & \checkmark & \checkmark & - & $1.58\pm2.73$ & $34.53\pm6.83$ \\ 
        \checkmark & \checkmark & \checkmark & \checkmark & $\pmb{1.56\pm2.52}$ & $\pmb{34.64\pm6.73}$ \\ 
        \bottomrule
    \end{tabular}
    }
    \tabmargin
\end{table}

\begin{table}[t]
    \centering
    \scriptsize
    \caption{
        \textbf{Analysis on alternatives of Hallucination-Net.}
        With the positive residual learning, the model predicts physically accurate values within the over-exposed regions.
        The resize convolution reduces the checkerboard artifacts, while the perceptual loss helps generate realistic details.
    }
    \label{tab:compare_hal}
    \begin{tabular}{ccc|c}
	    \toprule
	    Positive residual & Resize convolution & Perceptual loss & HDR-VDP-2 ($\uparrow$) \\
		\midrule
	    - & - & - & $63.60\pm15.32$ \\
	    \checkmark & - & - & $64.79\pm15.89$ \\
	    \checkmark & \checkmark & - & $64.52\pm16.05$ \\  
	    \checkmark & \checkmark & \checkmark & \pmb{$66.31\pm15.82$} \\
	    \bottomrule
    \end{tabular}
    \tabmargin
\end{table}

\heading{Hallucination.} 
We start with the architecture of Eilertsen~\etal~\cite{eilertsen2017hdrcnn}, which does not enforce the predicted residuals being positive.
As shown in~\tabref{compare_hal}, our model design (predicting positive residuals) can improve the performance by 1.19 HDR-VDP-2 scores.
By replacing the transposed convolution with the resize convolution in the decoder, our model effectively reduces the checkerboard artifacts.
Furthermore, introducing the perceptual loss for training not only improves the HDR-VDP-2 scores but also helps the model to predict more realistic details.
We provide visual comparisons in the supplementary material.

\heading{End-to-end training from scratch.}
To demonstrate the effectiveness of explicitly reversing the camera pipeline, we train our entire model (including all sub-networks) from scratch without any intermediate supervisions.
Compared to the proposed model shown in~\tabref{compare_stoa_hdrvdp}, the performance of such a model drops significantly (-4.43 and -3.48 HDR-VDP-2 scores in the \textsc{HDR-Synth} and \textsc{HDR-Real} datasets, respectively).
It shows that our stage-wise training is effective, and the performance improvement does not come from the increase of network capacity.

\section{Conclusions}
\label{sec:conclusion}

We have presented a novel method for single-image HDR reconstruction. 
Our key insight is to leverage the domain knowledge of the LDR image formation pipeline for designing network modules and learning to \emph{reverse} the imaging process.
Explicitly modeling the camera pipeline allows us to impose physical constraints for network training and therefore leads to improved generalization to unseen scenes.
Extensive experiments and comparisons validate the effectiveness of our approach to restore visually pleasing details for a wide variety of challenging scenes.  %

\vspace{1mm}
{
\heading{Acknowledgments.} This work is supported in part by NSF CAREER ($\#$1149783), NSF CRII ($\#$1755785), MOST 109-2634-F-002-032, MediaTek Inc. and gifts from Adobe, Toyota, Panasonic, Samsung, NEC, Verisk, and Nvidia.
}

\newpage
{\small
\bibliographystyle{ieee_fullname}
\bibliography{ldr2hdr}
}

\end{document}